\documentclass[conference]{IEEEtran}
\IEEEoverridecommandlockouts
\usepackage[utf8]{inputenc}
\usepackage{amsmath,graphicx,hyperref}
\usepackage{bm}
\usepackage[nolist]{acronym}
\usepackage{enumitem}
\usepackage{url}
\usepackage{graphicx}
\usepackage{colortbl} 
\usepackage{multirow}
\usepackage{xcolor}
\usepackage{textcomp}
\usepackage{dblfloatfix}    
\usepackage{lipsum}
\usepackage{comment}
\usepackage{hhline}
\usepackage{tabularx,ragged2e}
\usepackage{caption}
\usepackage{subcaption}
\usepackage{ragged2e}
\usepackage{esvect}
\usepackage{url}
\usepackage{algorithmic}
\usepackage{algorithm}
\usepackage{mathtools}
\usepackage{tikz,pgfplots}
\usetikzlibrary{calc}
\usepackage{tkz-euclide}
\pgfplotsset{compat=1.15}
\usepgfplotslibrary{fillbetween}
\usepackage[inkscapelatex=false]{svg}
\usepackage{amsfonts}
\usepackage[utf8]{inputenc}
\usepackage{anyfontsize}
\usepackage{optidef}
\usepackage{lipsum, babel}
\usepackage{booktabs}
\usepackage{amsthm}
\usepackage{tcolorbox}
\tcbuselibrary{theorems}
\usepackage{svg}

\newcolumntype{C}{>{\Centering\arraybackslash}X}

\newcommand{\rev}[1]{{\color{black}{#1}}}

\usepackage{calc}
\newlength\myheight
\newlength\mydepth
\settototalheight\myheight{Xygp}
\DeclareMathOperator*{\argmin}{arg\,min}

\ExplSyntaxOn
\NewDocumentCommand{\prow}{m}
{
	\seq_set_split:Nnn \l_tmpa_seq { , } { #1 }
	\seq_use:Nn \l_tmpa_seq { & }
	\\
}
\ExplSyntaxOff

\newcommand{\compl}{\mathbb{C}}      

\newcommand{\e}{{\rm e}} 

\newcommand{\expof}[1]{{\e}^{#1}}                
\newcommand{\normof}[2]{\left\|#1\right\|_{#2}}
\newcommand{\conj}{*}          
\newcommand{\trans}{{\rm T}}   
\newcommand{\herm}{{\rm H}}    

\newtcbtheorem{myExample}{Example}%
{colback=gray!5,colframe=lightgray!35!black,fonttitle=\bfseries}{th}
\newtcbtheorem{KD}{Key Difference}%
{colback=gray!5,colframe=lightgray!35!black,fonttitle=\bfseries}{th}

\title{Compressed Sensing-Driven Near-Field Localization Exploiting Array of Subarrays}
\author{
	\IEEEauthorblockN{ 
	   Sai Pavan Deram \IEEEauthorrefmark{1}\IEEEauthorrefmark{2},
       Jacopo Pegoraro \IEEEauthorrefmark{4},
       Javier Lorca Hernando  \IEEEauthorrefmark{3},
         Jesus O. Lacruz \IEEEauthorrefmark{2},
		 Joerg Widmer{\IEEEauthorrefmark{2}}   
	}
\IEEEauthorblockA{
	\IEEEauthorrefmark{1}
	Universidad Carlos III {Madrid},
        \IEEEauthorrefmark{2}
	IMDEA Networks Institute,
    \IEEEauthorrefmark{3}
    InterDigital Inc.,
        \IEEEauthorrefmark{4}
	University of Padova
}}

\begin{document}
%
\maketitle
\begin{acronym}
	\acro{SHARE}{Sparse Hierarchical Angle-Range Estimation}
	\acro{OMP}{Orthogonal Matching Pursuit}
	\acro{ISAC}{Integrated Sensing and Communication}
	\acro{CS}{Compressed sensing}
	\acro{RF}{radio frequency}
\end{acronym}

\begin{abstract}

   \rev{Near-field localization for ISAC
   requires large-aperture arrays, making fully-digital implementations prohibitively complex and costly. While sparse subarray architectures can reduce cost, they introduce severe estimation ambiguity from grating lobes. To address both issues, we propose SHARE (Sparse Hierarchical Angle-Range Estimation), a novel two-stage sparse recovery algorithm. SHARE operates in two stages. It first performs coarse, unambiguous angle estimation using individual subarrays to resolve the grating lobe ambiguity. It then leverages the full sparse aperture to perform a localized joint angle-range search. This hierarchical approach avoids an exhaustive and computationally intensive two-dimensional grid search while preserving the high resolution of the large aperture.
    Simulation results show that SHARE significantly outperforms conventional one-shot sparse recovery methods, such as Orthogonal Matching Pursuit (OMP), in both localization accuracy and robustness. Furthermore, we show that SHARE's overall localization accuracy is comparable to or even surpasses that of the fully-digital 2D-MUSIC algorithm, despite MUSIC having access to the complete, uncompressed data from every antenna element. SHARE therefore provides a practical path for high-resolution near-field ISAC systems.}
\end{abstract}
	

%
\section{Introduction}
\label{sec:intro}

Estimating source parameters is an important problem in array signal processing, with applications in wireless communications, radar, sonar, and many other fields. Sources are grouped into two categories according to their distance from the antenna array: far-field and near-field sources \cite{YZO+23} \cite{EMC+24}. As wireless systems move to higher frequencies like millimeter-wave and terahertz, and as antenna arrays grow larger, the near-field region around transceivers expands significantly \cite{ZBP24}.   This is especially important for integrated sensing and communication (ISAC) systems, where near-field effects make it possible to achieve precise localization and environmental sensing together with data transmission, which are key for future 6G networks and smart infrastructure.

Unlike far-field sources, which have approximately planar wavefronts, near-field spherical wavefronts encode distance information in the phase curvature of the received signals. This enables the simultaneous estimation of angle and range from a single anchor. To realize these benefits, previous studies extended classical high-resolution algorithms, such as MUSIC \cite{S86} and ESPRIT \cite{RK89}, from the far field to the near field. Specific extensions for joint angle and range estimation include near-field MUSIC using a 2D spectral search \cite{HB91}, as well as near-field versions of ESPRIT \cite{CS95} and polynomial root-finding methods \cite{WF93}. 
\rev{More recent works have demonstrated high-precision localization with massive MIMO testbeds \cite{SAS+22} and introduced machine learning (ML) to handle localization in complex propagation environments \cite{DeLima2023Machine}.}

\rev{However, these methods typically rely on a fully-digital array architecture, where each antenna element is connected to a dedicated radio frequency (RF) chain. This one-to-one mapping becomes prohibitively expensive and power-intensive for large arrays. To address this scalability issue, recent works have explored partitioning a large array into smaller subarrays. For instance, the algorithm in \cite{APLE} uses the differences in observed angles between subarrays, while the work in \cite{AOB24} explores modular designs to maintain a large aperture with fewer elements. While promising, these subarray-based strategies still generally require a dedicated RF chain per element. However, these strategies introduce a new challenge, as the physical separation of subarrays to create a sparse aperture results in grating lobes that cause severe ambiguity in parameter estimation.}

\rev{
To address these challenges, we propose a novel framework based on compressive sensing with sparse modular arrays. The main contributions of this work are as follows:
\begin{itemize}
    \item We introduce a low-complexity compressive sensing framework that simultaneously addresses the challenges of high hardware cost and the spatial ambiguity from grating lobes. Our model uses a small number of RF chains to process signals from a large number of antennas, enabling large-aperture systems at a fraction of the cost of fully-digital arrays.

    \item We introduce SHARE (Sparse Hierarchical Angle-Range Estimation), a novel two-stage algorithm specifically designed for this architecture. Stage~1 performs a robust, non-coherent spectrum combining for coarse angle estimation using dense subarrays to resolve ambiguities. Stage~2 then leverages the full sparse aperture for a high-resolution joint angle-range refinement.

    \item We demonstrate through extensive simulations that SHARE significantly outperforms both conventional one-shot compressive methods (2D-OMP) and classical fully-digital algorithms (2D-MUSIC) in challenging, multi-source near-field scenarios. We show that our approach is not only more accurate and robust but is also far more computationally efficient.
\end{itemize}
}

 {\bf{Notation.}} This paper employs {the} following mathematical notation: scalars are denoted by italic letters (e.g., $A$, $a$), vectors by bold lowercase letters (e.g., $\mathbf{a}$), and matrices by bold uppercase letters (e.g., $\mathbf{A}$). Key operators include the  matrix transpose ($.^\trans$),  conjugate ($.^\conj$), Hermitian transpose ($.^\herm$), Frobenius norm ($\normof{.}{F}$), and Moore-Penrose pseudoinverse ($.^\dagger$).
\section{System Model}
\label{sec:SystemModel}
In this section, we introduce the array configuration and develop the compressive signal model that forms the basis of our proposed localization framework. We begin by defining the physical array structure and the ideal near-field signal, then introduce the practical hybrid architecture and the resulting compressive measurement model.%
\subsection{Array Configuration}
We consider a linear array oriented along the x-axis, composed of $M$ omnidirectional antenna elements \rev{operating at a carrier frequency $f_c$ with corresponding wavelength $\lambda = c/f_c$.} The array is structured as a modular array of $P$ subarrays, with each subarray containing $M_0$ elements, such that $M = P M_0$. The intra-subarray element spacing is $d$, and the inter-subarray spacing between the first elements of adjacent subarrays is $d_p$. The global position of the $m$-th element in the $p$-th subarray in the $x,y,z$ space is given by the vector $\mathbf{s}_{p,m} = [(p-1)d_p + md,0, 0]^\trans$, where $m=0, \dots, M_0-1$ and $p=1, \dots, P$. The large overall aperture of this array, denoted by ${D = (P-1)d_p + (M_0-1)d},$ extends the near-field region, defined by the Rayleigh distance $Z_R = 2D^2/\lambda$, enabling high-resolution localization.

\begin{figure}[h]
    \centering
    \begin{subfigure}[b]{0.48\linewidth}
        \centering
        \includegraphics[width=\linewidth]{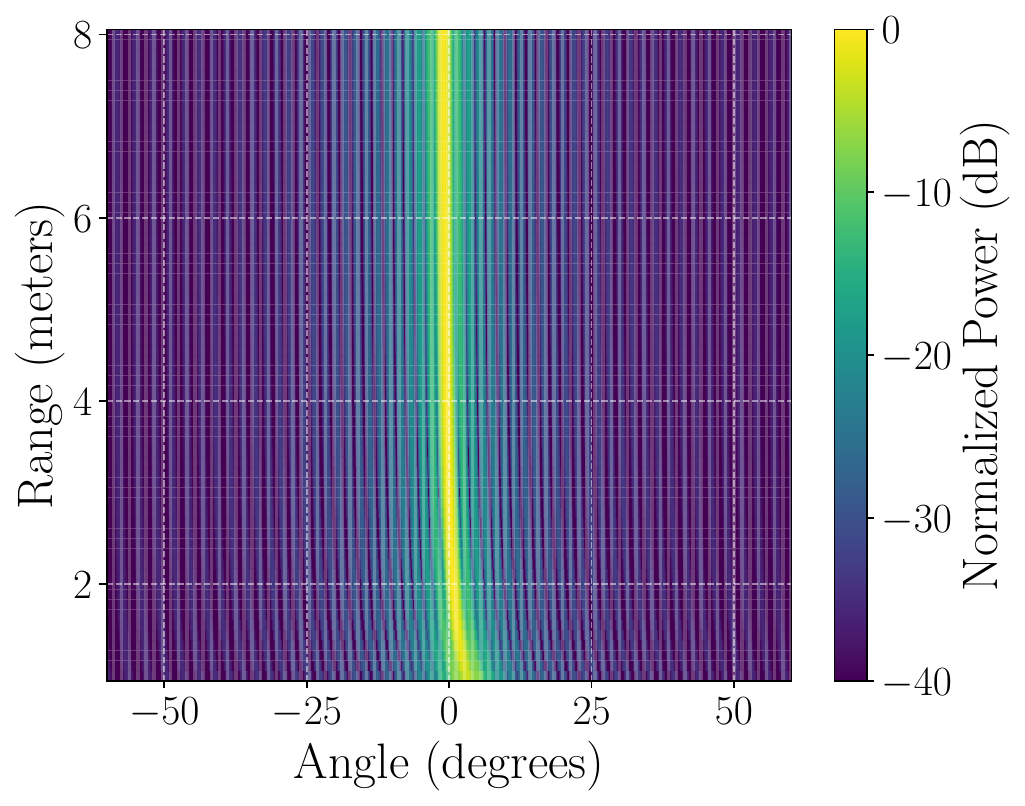}
        \caption{\rev{$d_p = M_0 d$}}
        \label{fig:Nograting} 
    \end{subfigure}
    \hfill 
    \begin{subfigure}[b]{0.48\linewidth}
        \centering
        \includegraphics[width=\linewidth]{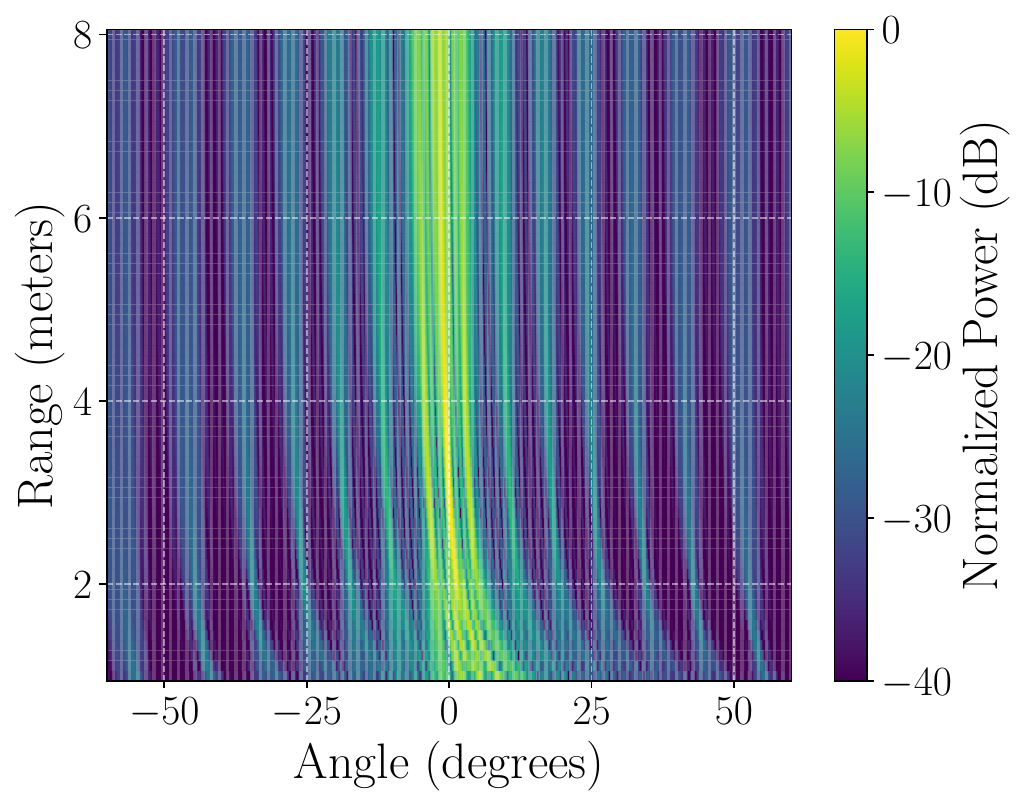}
        \caption{$d_p > M_0 d$}
        \label{fig:grating} 
    \end{subfigure}

    \caption{Comparison of the near-field beampattern for two subarray spacings. (a) When the spacing is small, a single unambiguous peak is formed. (b) When the spacing is large, ambiguous grating lobes appear.}
    \label{fig:beampattern}
\end{figure}

The inter-subarray spacing, $d_p$, is a critical design parameter that introduces a fundamental trade-off. \rev{When the subarrays are placed contiguously (i.e., $d_p=M_0 d$), uniformly sampled below the Nyquist spacing, its beampattern exhibits a single, unambiguous main lobe, as shown in Fig.~\ref{fig:Nograting}.} 
\rev{However, to maximize the aperture without increasing the element count, a sparse configuration with $d_p > M_0 d$ is often employed. In this case, the large, element-free gaps between subarrays mean the array is spatially undersampled across its full aperture.} This spatial undersampling introduces aliasing, which manifests as high-ambiguity grating lobes \footnote{While ``grating lobes'' technically refers to ambiguous lobes in a transmit beampattern, the same phenomenon of spatial aliasing creates parameter ambiguities in a receiver's spectral response. Due to reciprocity, we use the common and intuitive term ``grating lobes'' to refer to these ambiguities.} in the array response as shown in Fig.~\ref{fig:grating}. These ambiguities pose a significant challenge for conventional localization algorithms, making it difficult to determine a source's direction uniquely. Furthermore, such sparse configurations are often a practical necessity because physical constraints related to chip packaging and RF hardware can prevent subarrays from being placed immediately adjacent to one another.
\subsection{Fully Digital Array Signal Model}
Let us first consider an ideal fully-digital architecture where each of the $M$ antennas is connected to a dedicated radio frequency (RF) chain. We assume $L$ narrowband sources are located in the near-field, impinging on the array from the x-y plane. The location of the $l$-th source is denoted in spherical coordinates by its azimuth angle, $\theta_l$, and range, $r_l$, relative to the array's reference element ($\mathbf{s}_{1,0}$). The complex baseband signal received at the $(p,m)$-th antenna from all $L$ sources in the $n$-th temporal snapshot is
\begin{align}
	y_{p,m}(n) = \sum_{l=1}^{L} x_l(n) e^{-j2\pi f_c \tau_{p,m,l}} + v_{p,m}(n),
\end{align}
where $x_l(n)$ is the signal waveform transmitted by the $l$-th source, $f_c$ is the carrier frequency, and $v_{p,m}(n)$ is additive white Gaussian noise. The term $\tau_{p,m,l}$ represents the propagation delay of the $l$-th source signal relative to the reference element, given by
\begin{align}
	\tau_{p,m,l} = \frac{\|\mathbf{s}_{p,m} - \mathbf{s}_l\|_2 - r_l}{c},
\end{align}
$\mathbf{s}_l = r_l \boldsymbol{\kappa}_l (\theta_l, \phi_l)$ is the position vector of the $l$-th source, where $\boldsymbol{\kappa}_l (\theta_l,\phi_l) = \begin{bmatrix}
	\cos \theta_l \cos \phi_l, &\sin \theta_l \cos \phi_l, & \sin \phi_l
\end{bmatrix}^\trans$ is the direction vector and $c$ is the speed of light. The spherical nature of the wavefront is captured in this delay term, which is a non-linear function of the range, $r_l$ and angles $\theta_l, \phi_l$\footnote{As we consider a linear array on the x-axis, we assume all sources lie in the x-y plane, and thus set the elevation angle $\phi_l=0$ for simplicity, though the model can be readily extended to 2D planar arrays.} of the $l$-th source.

By stacking the received signals from all $M$ antennas over $N$ snapshots, we can express the signal model in a compact matrix form as:
\begin{equation*}
	\mathbf{Y} = \mathbf{A}\mathbf{X} + \mathbf{V},
	\label{eq:fully_digital_model}
\end{equation*}
where $\mathbf{Y} \in \mathbb{C}^{M \times N}$ is the received signal matrix, $\mathbf{X} \in \mathbb{C}^{L \times N}$ contains the source waveforms, and $\mathbf{V} \in \mathbb{C}^{M \times N}$ is the noise matrix. The matrix $\mathbf{A} \in \mathbb{C}^{M \times L}$ is the array manifold matrix, given by $\mathbf{A} = [\mathbf{a}(\theta_1, r_1), \dots, \mathbf{a}(\theta_L, r_L)]$, where each column is a near-field steering vector. The steering vector for a source at location $(\theta_l, r_l)$ is defined by the propagation delays to all $M$ antenna elements and $\forall p \in \{1,\ldots,P\}, m \in \{1,\ldots, M_0\}$
$$[\mathbf{a}(\theta_l, r_l)]_{(p,m)} = (\frac{r_l}{r_{p,m,l}})\expof{-j2\pi f_c \tau_{p,m,l}}.$$

While this model captures the complete spatial information, acquiring the full matrix $\mathbf{Y}$ necessitates $M$ dedicated RF chains, making the fully-digital approach impractical due to its prohibitive hardware cost and power consumption.
\subsection{Compressive Measurement Model}
\rev{To overcome the dual challenges of prohibitive hardware cost and the spatial ambiguity of sparse arrays, we adopt a low-complexity hybrid architecture that leverages compressive sensing (CS). In this framework, the $M_0$ antenna elements within each subarray are connected to a single RF chain via an analog combining network, reducing the total number of RF chains from $M$ to $P$. To capture sufficient spatial information, we collect $K$ distinct measurement snapshots by varying the settings of the analog phase shifters over time. Each snapshot provides a unique linear projection of the high-dimensional spatial signal. As we will show, this process not only enables parameter recovery with fewer RF chains but also helps mitigate the grating lobe problem by transforming sharp ambiguous peaks into a lower, more uniform sidelobe floor.}

At each measurement instant $k \in \{1, \dots, K\}$, the analog combining network for the entire array is described by a matrix $\mathbf{\Phi}_k \in \mathbb{C}^{P \times M}$ with a block-diagonal structure:
\begin{equation} \label{eqn:Phik}
	\mathbf{\Phi}_k = 
	\begin{bmatrix}
		\mathbf{w}_{k,1}^\trans & \mathbf{0}_{M_0}^\trans & \dots & \mathbf{0}_{M_0}^\trans \\
		\mathbf{0}_{M_0}^\trans & \mathbf{w}_{k,2}^\trans & \dots & \mathbf{0}_{M_0}^\trans \\
		\vdots & \vdots & \ddots & \vdots \\
		\mathbf{0}_{M_0}^\trans & \mathbf{0}_{M_0}^\trans & \dots & \mathbf{w}_{k,P}^\trans
	\end{bmatrix},
\end{equation}
where $\mathbf{w}_{k,p} \in \mathbb{C}^{M_0}$ is the vector of complex weights (e.g., $w_{k,p,m} = \alpha_{k,p,m} \,e^{j\psi_{k,p,m}}$) applied by the phase shifters in the $p$-th subarray. {These weights are typically implemented with a constant-modulus constraint, i.e., $|w_{k,p,m}|=1$, reflecting the use of analog phase shifters.}

\begin{figure}[t]
    \centering 
    \begin{subfigure}[b]{0.48\linewidth}
        \centering
        \includegraphics[width=\linewidth]{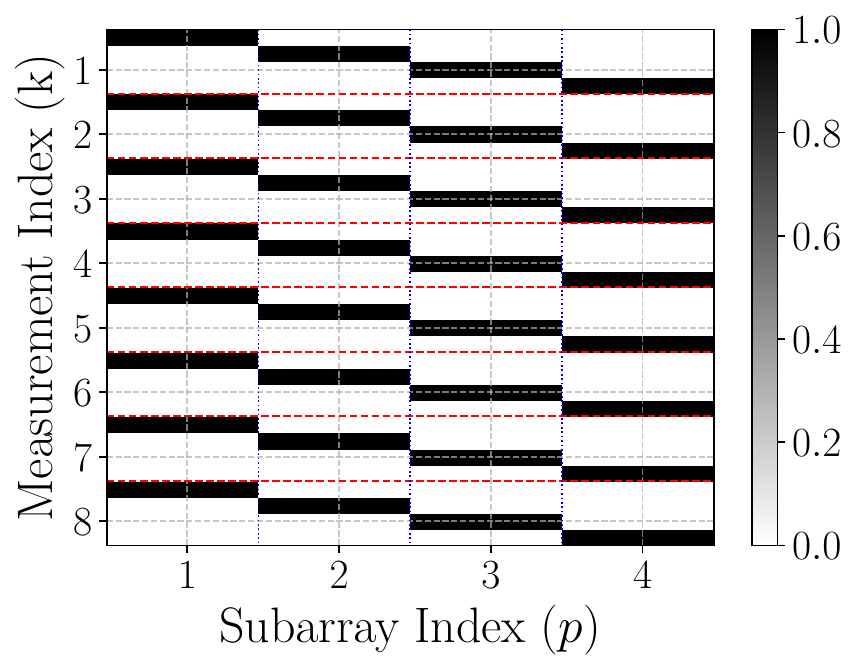}
        \caption{}
        \label{fig:phi_structure} 
    \end{subfigure}
    \hfill 
    \begin{subfigure}[b]{0.48\linewidth}
        \centering
        \includegraphics[width=\linewidth]{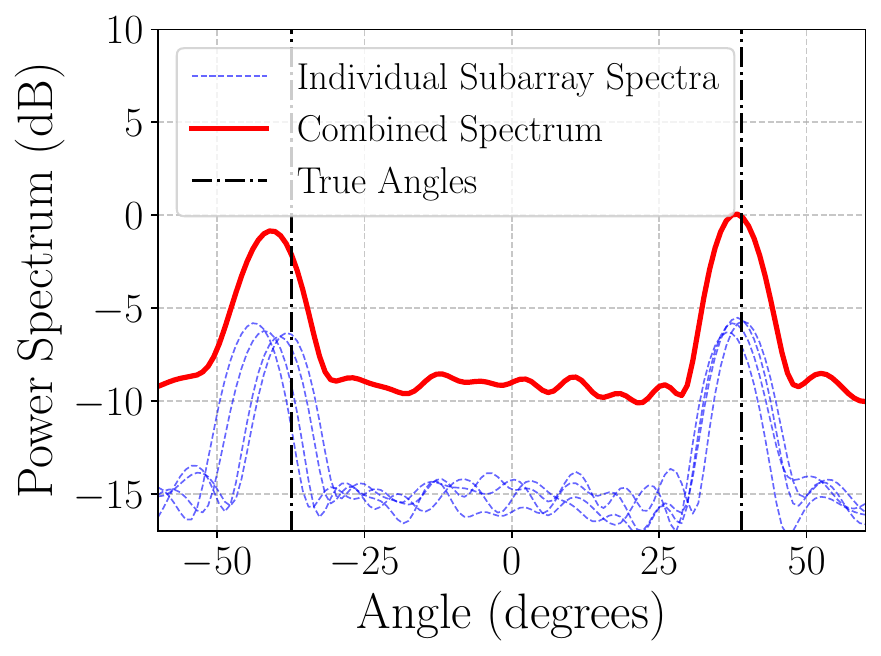}
        \caption{}
        \label{fig:stage1_spectrum} 
    \end{subfigure}
    \caption{Visualization of the proposed framework. (a) The block-diagonal, stacked structure of the compressive measurement matrix $\mathbf{\Phi}$. (b) The non-coherent combining of individual subarray spectra (blue) yields a robust combined spectrum (red) with clear peaks at the true source angles (dot-dashed black).}
    \label{fig:framework_visualization} 
\end{figure}

By stacking the measurements from all $K$ instants, we form the overall compressive measurement matrix $\mathbf{\Phi} = [\mathbf{\Phi}_1^\trans, \dots, \mathbf{\Phi}_K^\trans]^\trans \in \mathbb{C}^{PK \times M}$. \rev{The structure of this matrix, which visually represents our measurement protocol, is illustrated in Fig.~\ref{fig:phi_structure}.} The final measurement model is a Multiple Measurement Vector (MMV) problem \cite{YANG2018509}, as it jointly processes all $N$ temporal snapshots which share a common sparsity pattern determined by the source locations. The model is given by
\begin{equation}
	\tilde{\mathbf{Y}} = \mathbf{\Phi}\mathbf{Y} = \mathbf{\Phi}(\mathbf{A}\mathbf{X} + \mathbf{V}) = \mathbf{\Phi}\mathbf{A}\mathbf{X} + \mathbf{Z},
	\label{eq:cs_model_final}
\end{equation}
where $\tilde{\mathbf{Y}} \in \mathbb{C}^{PK \times N}$ is the compressed data available for processing and $\mathbf{Z} = \mathbf{\Phi}\mathbf{V}$ is the effective noise. Assuming the original noise elements in $\mathbf{V}$ are i.i.d. with variance $\sigma_v^2$, the effective noise $\mathbf{Z}$ is still zero-mean Gaussian, but its spatial covariance is now correlated and given by $\sigma_v^2 \mathbf{\Phi}\mathbf{\Phi}\herm$. Our objective is to estimate the source parameters $\{\theta_l, r_l\}_{l=1}^L$ from the low-dimensional measurements $\tilde{\mathbf{Y}}$, thus enabling high-resolution localization with significantly reduced hardware.


\section{SHARE: A Hierarchical Estimation Algorithm}
The problem of jointly estimating the parameters of $L$ sources can be formulated as a non-convex optimization problem. The goal is to find the set of angles $\bm{\Theta} = \{\theta_1, \dots, \theta_L\}$, ranges $\mathbf{R} = \{r_1, \dots, r_L\}$, and the signal waveform matrix $\mathbf{X} \in \mathbb{C}^{L \times N}$ that best explain the measurements:
\begin{equation}
    \argmin_{\bm{\Theta}, \mathbf{R}, \mathbf{X}}  \left\| \tilde{\mathbf{Y}} - \mathbf{\Phi} \mathbf{A}(\bm{\Theta}, \mathbf{R}) \mathbf{X} \right\|_F^2 ,
    \label{eq:joint_problem_multi}
\end{equation}
where $\mathbf{A}(\bm{\Theta}, \mathbf{R})$ is the array manifold matrix corresponding to the source locations. While a direct solution to \eqref{eq:joint_problem_multi} is possible by applying a greedy algorithm like OMP over a fine-grained two-dimensional grid, this brute-force approach is computationally prohibitive. To overcome this, we propose the Sparse Hierarchical Angle-Range Estimation (SHARE) algorithm, a two-stage procedure that decouples the estimation problem to achieve both computational efficiency and high accuracy.

The first stage of SHARE aims to find an unambiguous, albeit coarse, estimate of the source angles by leveraging the properties of individual subarrays. For any single subarray, the aperture $(M_0-1)d$ is small, ensuring two key properties: 1)~the subarray is dense ($d \le \lambda/2$) and therefore free of grating lobe ambiguities, and 2)~sources are likely to be in the subarray's far-field, i.e., $r \gg 2((M_0-1)d)^2/\lambda$. This second property allows us to accurately approximate the near-field steering vector with its simpler far-field equivalent. To show this, let us re-examine the propagation delay. Under the far-field condition ($r_l \gg \|\mathbf{s}_{p,m}\|_2$), the Fresnel approximation for the source distance $\|\mathbf{s}_{p,m} - \mathbf{s}_l\|_2$ is:
\begin{align}
    \|\mathbf{s}_{p,m} - \mathbf{s}_l\|_2 \approx r_l - \mathbf{s}_{p,m}^\trans \boldsymbol{\kappa}_l + \frac{\|\mathbf{s}_{p,m}\|_2^2}{2r_l} - \dots
\end{align}
The term $r_l - \mathbf{s}_{p,m}^\trans \boldsymbol{\kappa}_l$ represents the standard planar wave approximation. The term $\frac{\|\mathbf{s}_{p,m}\|_2^2}{2r_l}$ contains the spherical wavefront curvature information, which is negligible for a small-aperture subarray. The delay can thus be accurately approximated by the first-order term, which for an element at position $md$ on the x-axis simplifies to $\tau_{m,l} \approx -\frac{md \cos\theta_l}{c}$. Substituting this into the steering vector definition yields the range-independent far-field model:
\begin{equation}
    [\mathbf{a}_{\text{sub}}(\theta)]_m = e^{-j2\pi f_c \tau_{m,l}} \approx e^{j\frac{2\pi}{\lambda} (md \cos\theta)}.
\end{equation}

\rev{First, we partition the full measurement matrix in \eqref{eq:cs_model_final} into $P$ blocks, $\tilde{\mathbf{Y}}_p \in \compl^{K \times N}$, where each block contains the measurements corresponding to the $p$-th subarray. To obtain a robust coarse estimate, we leverage information from all $P$ subarrays. A theoretically optimal approach would be to coherently combine the measurements before the spectral search. This would involve compensating for the phase offsets between subarrays, which are a function of the subarray spacing $d_p$ and the unknown angle $\theta$. The coherently combined complex spectrum vector, $\mathbf{b}_{\text{c}}(\theta)$, would be formed as:
\begin{equation}
    \mathbf{b}_{\text{c}}(\theta) = \sum_{p=1}^{P} \left( \mathbf{a}_{\text{sub}}(\theta)^\herm \mathbf{\Phi}_p^\herm \tilde{\mathbf{Y}}_{p}  e^{-j\frac{2\pi}{\lambda} (p-1)d_p \cos\theta} \right).
    \label{eq:coherent_combining}
\end{equation}
However, implementing \eqref{eq:coherent_combining} is impractical, as it requires knowing the true angle $\theta$ to calculate the phase compensation term needed to search for that same angle, creating a circular dependency. This method would also re-introduce the grating lobe ambiguities from the full sparse array.

Therefore, to create a robust and practical estimator, we employ non-coherent combining, which operates on the signal power, thereby discarding the problematic phase information. First, we compute the individual power spectrum for each subarray $p$ by projecting its measurements onto the far-field steering vectors:
\begin{equation}
    \mathcal{P}_p(\theta) = \|\mathbf{a}_{\text{sub}}(\theta)^\herm \mathbf{\Phi}_p^\herm \tilde{\mathbf{Y}}_{p}\|_F^2.
\end{equation}
As this power calculation is independent of the inter-subarray phase offsets, each $\mathcal{P}_p(\theta)$ provides an unambiguous but potentially noisy estimate of the angular spectrum. We can then robustly combine these real-valued power spectra through summation:
\begin{equation}
	\mathcal{P}_{\text{total}}(\theta) = \sum_{p=1}^{P} \mathcal{P}_p(\theta).
\end{equation}
This summation acts as an averaging process, enhancing the peaks corresponding to true sources while suppressing noise and spurious artifacts. The $L$ largest peaks of this combined spectrum, $\mathcal{P}_{\text{total}}(\theta)$, provide the set of coarse angle estimates, $\hat{\Theta}_0 = \{\hat{\theta}_1, \dots, \hat{\theta}_L\}$.}
With the set of coarse angle estimates $\hat{\Theta}_0$ obtained, the second stage of SHARE performs a high-resolution joint refinement of both angle and range. For clarity, we first develop the procedure for a single source ($L=1$). We assume our measurement matrix, now denoted $\tilde{\mathbf{Y}}_1 \in \mathbb{C}^{PK \times N}$, contains one source with signal waveform $\mathbf{x}_1 \in \mathbb{C}^{1 \times N}$ at location $(\theta, r)$. The corresponding measurement model is:
\begin{equation}
	\tilde{\mathbf{Y}}_1 = \mathbf{\Phi} \mathbf{a}(\theta, r) \mathbf{x}_1 + \mathbf{Z}_1.
	\label{eq:single_source_model_stage2}
\end{equation}
Given the coarse angle estimate $\theta_0$ from Stage~1, the true angle is $\theta = \theta_0 + \delta$, where $\delta$ is a small error. The estimation problem for this single source is to find the angular correction $\delta$, range $r$, and signal waveform $\mathbf{x}_1$ that best explain the data:
\begin{equation}
	\argmin_{\delta, r, \mathbf{x}_1} \left\| \tilde{\mathbf{Y}}_1 - \mathbf{\Phi} \mathbf{a}(\theta_0 + \delta, r) \mathbf{x}_1 \right\|_F^2.
	\label{eq:joint_refine_problem}
\end{equation}
The objective in \eqref{eq:joint_refine_problem} is non-convex. However, since the optimal signal waveform $\mathbf{x}_1$ for any given location $(\theta, r)$ can be found in a least-squares sense, the problem can be reduced to a joint search over only the two continuous location parameters, $\delta$ and $r$. This remaining problem is still challenging, and a direct solution would typically require iterative optimization methods.

To create a computationally efficient and non-iterative solution, we instead transform this continuous search into an equivalent discrete sparse recovery problem. We construct a fine-grained 2D search grid in the neighborhood of the coarse estimate $\theta_0$. A refined angle grid, $\Theta_{\text{ref}}$, is created by sampling a small window spanning $\pm\Delta_\theta$ with a step size of $\delta_\theta$, containing $G_\delta = 2\Delta_\theta/\delta_\theta + 1$ points. Similarly, the range grid, $\mathcal{R}_{\text{grid}}$, is defined with $G_r$ distinct points. Based on this localized grid, we build a refined dictionary matrix $\mathbf{\Psi} \in \mathbb{C}^{PK \times G}$:
\begin{equation}
	\mathbf{\Psi} = [\dots, \mathbf{\Phi}\mathbf{a}(\theta_i, r_j), \dots], \quad \forall \theta_i \in \Theta_{\text{ref}}, r_j \in \mathcal{R}_{\text{grid}}.
\end{equation}
The cardinality of this dictionary for a single source is $G = G_\delta  G_r$. The estimation problem can now be cast as a sparse recovery problem to find a sparse matrix $\mathbf{S} \in \mathbb{C}^{G \times N}$ with one non-zero row:
\begin{equation}\label{eq:opt-prob-single}
	\min_{\mathbf{S}} \|\mathbf{S}\|_{2,0} \quad \text{s.t.} \quad \|\tilde{\mathbf{Y}}_1 - \mathbf{\Psi}\mathbf{S}\|_F \leq \epsilon,
\end{equation}
where $\epsilon$ is a noise-dependent parameter.
To generalize this to the full multi-source problem, we construct the dictionary $\mathbf{\Psi}$ to cover the neighborhoods of all $L$ coarse estimates from Stage~1. This is done by taking the union of the individual search windows around each coarse angle, i.e., $\Theta_{\text{ref}} = \bigcup_{l=1}^{L} \Theta_{\text{ref}, l}$. The cardinality of this unified dictionary is approximately $G \approx L  G_\delta  G_r$, which remains significantly smaller than that of a full global grid. The estimation problem then takes the same form as in \eqref{eq:opt-prob-single}, but we now use the full measurement matrix $\tilde{\mathbf{Y}}$ and solve for an $L$-sparse matrix $\mathbf{S}$ whose non-zero rows correspond to the $L$ source locations.

This hierarchical procedure effectively solves the overall optimization problem \eqref{eq:joint_problem_multi}. Stage~1 provides the coarse angles $\hat{\Theta}_0$ which guide the construction of the localized dictionary. Stage~2 then solves for the refined angle and range by finding the sparsest solution using a greedy algorithm such as Orthogonal Matching Pursuit (OMP). The corresponding non-zero rows of the estimated $\mathbf{S}$ then implicitly provide the signal waveform estimate via a final least-squares projection. This two-stage approach efficiently finds the three unknown parameters with significantly reduced computational complexity.

\section{Simulation Results}
In this section, we evaluate the performance of the proposed SHARE algorithm through extensive Monte Carlo simulations. We compare it against two primary baselines: the one-shot 2D-OMP algorithm \cite{NFOMP}, which operates on the same compressive data, and the classical 2D-MUSIC algorithm \rev{adopted for near-field }\cite{NFmusic}, which assumes a fully-digital array.

The primary performance metric is the Root Mean Squared Error (RMSE), which quantifies the average deviation between an estimated parameter vector $\hat{\bm{\mu}}$ and its true value $\bm{\mu}$. It is computed as $\text{RMSE} = \sqrt{\frac{1}{L} \sum_{l=1}^L \|\hat{\bm{\mu}}_{(l)} - \bm{\mu}_l\|^2 }$.
The RMSE is then averaged over multiple Monte Carlo realizations, where the parameter $\bm{\mu}$ represents the angle $\theta$, range $r$, or the 3D position vector $\mathbf{s}$. The position vector is inferred from the estimated angle and range via the transformation $\hat{\mathbf{s}} = [\hat{r}\cos(\hat{\theta}), \hat{r}\sin(\hat{\theta}), 0]^\trans$.

Unless stated otherwise, all results are averaged over 2500 independent realizations. We simulate a sparse modular array with $P=4$ subarrays and $M_0=16$ elements each, operating at $f_c = 60.48$~GHz. The intra-subarray spacing is $d = \lambda/2$, and the inter-subarray spacing is $d_p = 16\lambda$ \cite{MIMORPH}. We use $N=32$ temporal snapshots and $K=16$ compressive measurements, with the combining vectors chosen from a $M_0 \times M_0$ DFT matrix. A key aspect of our comparison is the search grid: the benchmark algorithms utilize a fine-grained global grid with $G_\theta =121$ angular points within the interval $[-60^\circ, 60^\circ]$ and $G_r = 64$ range points in the interval $[1, 9]\,\text{m}$. In contrast, SHARE uses a coarse $G_{\theta_c}=41$-point angular grid in Stage~1, and its refined Stage~2 search is performed over a small local grid within a $\Delta_{\theta} = \pm3^\circ$ window around each coarse estimate.

\begin{figure}[htbp]
    \centering 
    \includegraphics[width=\linewidth]{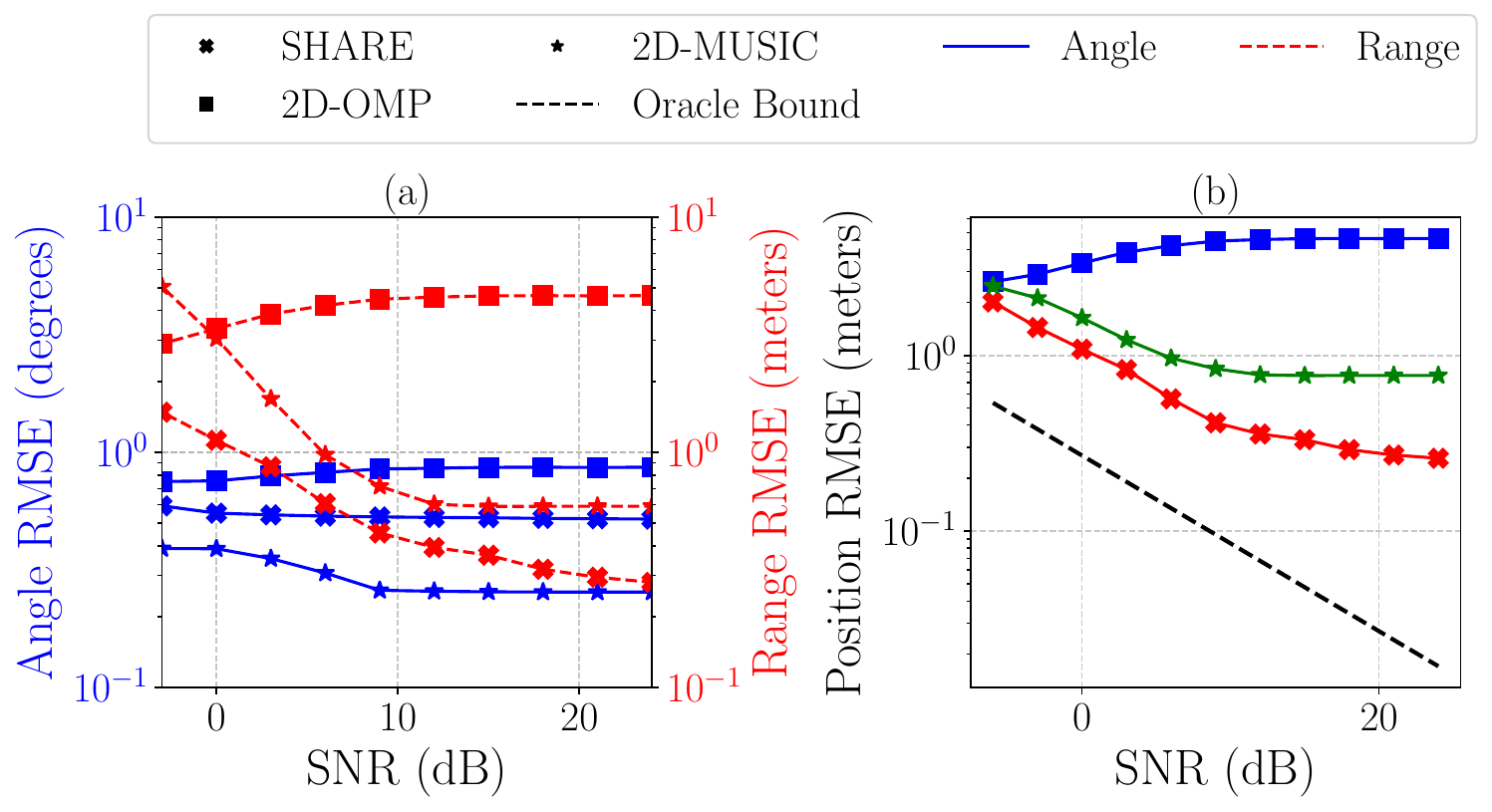}
    \caption{Performance comparison for two closely spaced sources located at ($43.3^\circ, 4.8\,\text{m}$) and ($43.8^\circ, 4.6\,\text{m}$). (a) Joint Angle and Range RMSE vs. SNR. (b) Overall Position RMSE vs. SNR.}
    \label{fig:rmse_closeledspaced}
\end{figure}

In the first experiment, we evaluate performance in a challenging scenario with two closely spaced near-field sources, located at ($43.3^\circ, 4.8\,\text{m}$) and ($43.8^\circ, 4.6\,\text{m}$). Figure~\ref{fig:rmse_closeledspaced} presents the mean Root Mean Squared Error (RMSE) versus SNR.
As shown in Fig.~\ref{fig:rmse_closeledspaced}(a), the one-shot 2D-OMP algorithm exhibits a high error floor for both angle and range. This indicates a failure to resolve the two sources due to dictionary coherence. The fully-digital 2D-MUSIC algorithm demonstrates a notable trade-off, achieving high-precision super-resolution performance in the angle domain but failing to estimate the range.
In contrast, the proposed SHARE algorithm shows steadily improving performance with SNR in both domains. Its angle accuracy approaches that of MUSIC, while its range RMSE is the only one that decreases significantly. This demonstrates its ability to leverage the full sparse aperture for range estimation after resolving the initial ambiguity.

The overall position RMSE, which combines both angle and range errors, is shown in Fig.~\ref{fig:rmse_closeledspaced}(b). The trends in this plot are a direct consequence of the individual parameter performance. The high error floors for 2D-OMP and 2D-MUSIC are dominated by their respective failures in resolving the sources and estimating the range. The SHARE algorithm's performance is again superior, with its accuracy consistently trending towards the Oracle bound. A slight flattening of its curve is observed at high SNRs, which indicates that its performance becomes limited by the resolution of the search grid used in Stage~2.

\begin{figure}[tbp]
    \centering 
    \includegraphics[width=\linewidth]{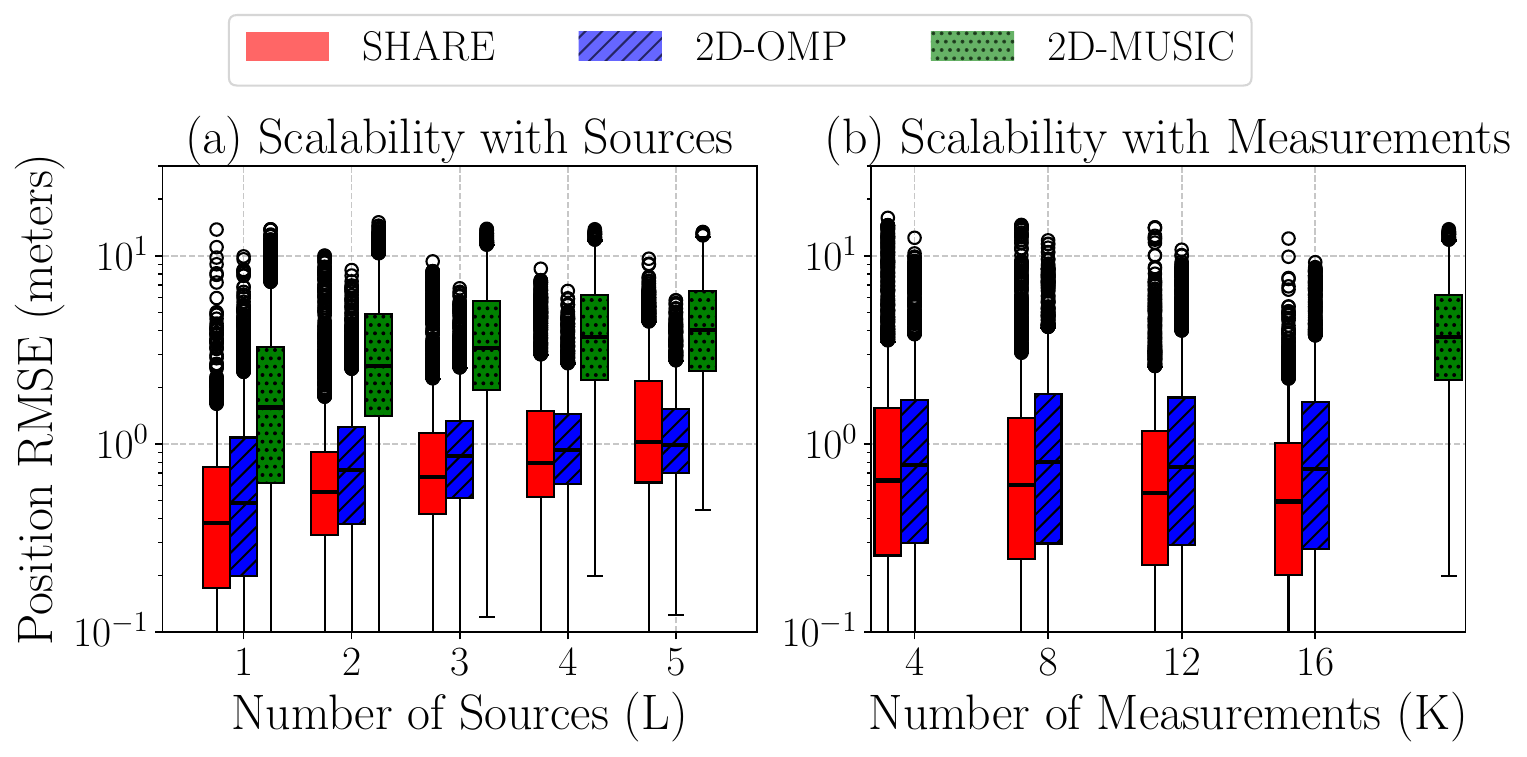}
    \caption{Algorithm scalability analysis with random source parameters at a fixed SNR of 10~dB. (a) Position RMSE versus the number of sources. (b) Position RMSE versus the number of compressive measurements.}
    \label{fig:scalability_analysis}
\end{figure}

\rev{To analyze the robustness and scalability of the algorithms, we plot the full distribution of the position RMSE in Figure~\ref{fig:scalability_analysis}. These results are for randomly located sources at a fixed SNR of 10~dB, with source locations drawn from a uniform distribution spanning an angular range of $[-60^\circ, 60^\circ]$ and a distance of $[1, 10]\,\text{m}$.

The box plots in Figure~\ref{fig:scalability_analysis}(a) show the performance as the number of sources ($L$) increases. We observe that the proposed SHARE algorithm consistently achieves the lowest median RMSE (the black line within the box), indicating that its typical performance is superior to the baselines. As $L$ increases, all algorithms exhibit performance degradation. The plot also reveals that in these challenging random multi-source scenarios, all three methods are susceptible to a significant number of failures, evidenced by the number of outliers. While SHARE's median error degrades gracefully, its primary advantage over 2D-OMP in this scenario is its accuracy.

Figure~\ref{fig:scalability_analysis}(b) highlights a key advantage of SHARE by showing its performance as a function of the number of compressive measurements ($K$). We observe that SHARE's median error and overall distribution consistently improve as more measurements are provided. This indicates that our hierarchical algorithm effectively utilizes the additional data to refine its estimates. In contrast, the performance of 2D-OMP is largely stagnant, showing no significant improvement with more measurements. This suggests its performance is limited by the dictionary structure rather than a lack of information. This analysis confirms that SHARE is not only more accurate on average but also makes far more efficient use of the compressive measurements, providing a more scalable solution for practical localization systems.}

\subsection*{Computational Complexity Analysis}
We analyze the computational complexity of SHARE and compare it against the fully-digital 2D-MUSIC and the one-shot compressive 2D-OMP algorithms. The analysis focuses on the dominant computational steps. Let the size of the full angle and range search grids be $G_\theta$ and $G_r$, respectively. For SHARE, the coarse angle grid is of size $G_{\theta_c}$ and the refined local angle grid is of size $G_{\delta}$.

The complexity of 2D-MUSIC is dominated by the eigendecomposition of the $M \times M$ covariance matrix, at $\mathcal{O}(M^3)$, and the 2D spectral search over all $G_\theta G_r$ grid points, costing $\mathcal{O}(G_\theta G_r M^2)$. The one-shot 2D-OMP algorithm performs $L$ iterations, where the dominant cost in each is correlating all $G_\theta G_r$ dictionary atoms with the $PK \times N$ residual matrix, resulting in a total complexity of $\mathcal{O}(L  PKN  G_\theta G_r)$.

The proposed SHARE algorithm achieves its efficiency by decoupling this search into two much smaller stages. In Stage 1, the dominant online task is computing the 1D power spectrum for each of the $P$ subarrays over the coarse angle grid. This involves correlating the pre-computed subarray dictionary with the $K \times N$ measurement block, for a total complexity of $\mathcal{O}(P N K G_{\theta_c})$. Subsequently, Stage 2 performs an OMP search over a drastically reduced 2D grid of size $G_{\delta} G_r$. The complexity of this localized refinement is $\mathcal{O}(L  PKN  G_{\delta} G_r)$. The total complexity of SHARE is therefore the sum of these two stages. By transforming the burdensome multiplicative search over $G_\theta G_r$ into an additive combination of two smaller searches, SHARE achieves a substantial reduction in computational load, especially for fine grid resolutions. A summary is provided in Table~\ref{tab:complexity}. For the simulation parameters provided below Table \ref{tab:complexity}, SHARE is approximately 8 times efficient than OMP while being more robust and accurate.

\begin{table}[htbp]
\centering
\caption{Computational Complexity Comparison}
\label{tab:complexity}
\renewcommand{\arraystretch}{1.5} 
\begin{tabular}{llc}
\hline
\textbf{Algorithm} & \textbf{ Complexity} & \textbf{FLOPS} \\
\hline
\begin{minipage}[t]{2cm}
    2D-MUSIC (FD)
\end{minipage}
    & \begin{minipage}[t]{4cm}
        $\mathcal{O}(M^3 + G_\theta G_r M^2)$
    \end{minipage}  
    & \begin{minipage}[t]{1.5cm}
       $\sim 3.19 \times 10^7$
    \end{minipage}   \\
\begin{minipage}[t]{2cm}
    2D-OMP (CS)
\end{minipage}
    & \begin{minipage}[t]{4cm}
       $\mathcal{O}(L  PKN  G_\theta G_r)$
    \end{minipage}  
    & \begin{minipage}[t]{1.5cm}
      $\sim 1.58 \times 10^{7}$ 
    \end{minipage}  \\
\begin{minipage}[t]{2cm}
   \textbf{SHARE (CS)}
\end{minipage}
    & \begin{minipage}[t]{4cm}
      $\mathcal{O}(P N K G_{\theta_c}+LPKN  G_{\delta} G_r)$
    \end{minipage}  
    & \begin{minipage}[t]{1.5cm}
        $\sim 1.91 \times 10^6$ 
    \end{minipage} \\
\hline
\end{tabular}
\newline 
	\footnotesize
\begin{minipage}{\columnwidth}
\vspace{0.5em} 
    \raggedright
    {$M$: Sensors, $N$: Snapshots, $L$: sources, $P$: RF chains, $M_0$: sensors in each subarray, $K$: compressive measurements, $G_{\theta}$: angular grid size, $G_{r}$: range grid size, $G_{\theta_{c}}$: coarse theta grid size, $G_{\delta}$: localized grid size}
    \scriptsize{}
\end{minipage}
\begin{minipage}{\columnwidth}
    \footnotesize 
     $M=64,~ P=4,~ M_0=16, ~L=1,~ N=32,~ K=16,~ G_\theta=121,~ G_r=64,~ G_{\theta_c}= 41, ~G_{\delta}=14$
\end{minipage}
\end{table}
\vspace{-3mm}

\section{Conclusions}\label{sec:conclusions}
In this paper, we have introduced SHARE, a hierarchical algorithm that enables practical high resolution near field localization using low cost compressive subarray architectures. We have shown that by strategically decoupling the estimation problem, it is possible to overcome the limitations of conventional methods. The algorithm first resolves angular ambiguities with small dense subarrays and then refines both angle and range with the full sparse aperture. From our results, we have demonstrated that SHARE is significantly more computationally efficient, accurate, and robust than both brute force compressive sensing and fully digital benchmark algorithms, especially in challenging multi-source scenarios. As directions for future work, we plan to investigate the design of optimized sensing matrices to further enhance estimation accuracy. Furthermore, we intend to validate the proposed algorithm on a practical hardware testbed to assess its performance in real-world operating conditions. In such practical implementations, incorporating mutual coupling effects into the signal model will be essential to compensate for hardware non-idealities and ensure the algorithm's efficiency.

\section*{Acknowledgment}
This work received funding from the EU's Horizon Europe program
under Grant Agreement No. 101192521
(MultiX), and the Madrid Regional Government
through Projects TUCAN6-CM (TEC-2024/COM-460) and DISCO6G-CM (TEC-2024/
COM-360).


\bibliographystyle{IEEEtran}
\bibliography{ICASSP2026/references/refs}

\end{document}